\documentclass{aa}  
\usepackage{booktabs}
\usepackage{threeparttable}
\usepackage{graphicx}
\usepackage{txfonts}
\begin{document}


\title{Stellar mass - halo mass relation for the brightest central galaxies of X-ray clusters since $z\sim0.65$ }
\titlerunning{Stellar mass - halo mass relation since $z\sim0.65$}
\authorrunning{G.Erfanianfar}

\author{G.~Erfanianfar\inst{1}\and A.~Finoguenov \inst{1,2} \and K.~Furnell\inst{3} \and P.~Popesso\inst{4} \and A.~Biviano\inst{5} \and S.~Wuyts\inst{6} \and C.~A.~Collins\inst{3}  \and M.~Mirkazemi\inst{4} \and J.~Comparat\inst{1} \and H.~Khosroshahi \inst{7} \and K. Nandra\inst{1} \and R.~Capasso\inst{8,4} \and E.~Rykoff\inst{9,10}   \and D.~Wilman\inst{1} \and A.~Merloni\inst{1} \and N.~Clerc\inst{11} \and M.~Salvato\inst{1}\and  J.~I.~Chitham\inst{1} \and \ L.~S.~Kelvin\inst{3}  \and G.~Gozaliasl\inst{12,2,13} \and A.~Weijmans\inst{14}  \and J.~Brownstein\inst{15} \and  E.~Egami\inst{16}  \and M.~J.~Pereira\inst{16} \and D.~P.~Schneider\inst{17,18} \and C.~ Kirkpatrick\inst{2} \and S.~Damsted \inst{2}\and A.~Kukkola\inst{2}}

\institute{Max planck institute for extraterrestrial physics, Garching by Munich, Germany
\and
University of Helsinki, Department of Physics, P.O. Box 64, FI-00014 Helsinki
\and
Astrophysics Research Institute, Liverpool John Moores University, IC2, Liverpool Science Park, 146 Brownlow Hill, Liverpool, L3 5RF, UK
\and
Excellence Cluster Universe, Boltzmannstr. 2, 85748 Garching bei M\"{u}nchen, Germany
\and
INAF/Osservatorio Astronomico di Trieste, via G. B. Tiepolo 11, 34131 Trieste, Italy
\and
Department of Physics, University of Bath, Claverton Down, Bath, BA2 7AY, UK 
\and
School of Astronomy, Institute for Research in Fundamental Sciences (IPM), Tehran, 19395-5531 Iran
\and
Faculty of Physics, Ludwig-Maximilians-Universit{\"a}t, Scheinerstr. 1, 81679 Munich, Germany
\and
Kavli Institute for Particle Astrophysics \& Cosmology, P. O. Box 2450, Stanford University, Stanford, CA 94305, USA
\and
SLAC National Accelerator Laboratory, Menlo Park, CA 94025, USA
\and
IRAP, Universit{\`a} de Toulouse, CNRS, UPS, CNES, Toulouse, France
\and
Finnish centre for Astronomy with ESO (FINCA), Quantum, Vesilinnantie 5, University of Turku, FI-20014 Turku, Finland
\and
Helsinki Institute of Physics, University of Helsinki, P.O. Box 64, FI-00014, Helsinki, Finland
\and
School of Physics and Astronomy, University of St Andrews, North Haugh, St Andrews, KY16 9SS, UK
\and
Department of Physics and Astronomy, University of Utah, 115 S. 1400 E., Salt Lake City, UT 84112, USA
\and
Steward Observatory, University of Arizona, 933 North Cherry Avenue, Tucson, AZ 85721, USA
\and
 Department of Astronomy and Astrophysics, The Pennsylvania State University, University Park, PA 16802
\and
 Institute for Gravitation and the Cosmos, The Pennsylvania State University, University Park, PA 16802
}

\date{ Accepted}

\label{firstpage}

\abstract{We present the brightest cluster galaxies (BCGs) catalog for SPectroscoic IDentification of eROSITA Sources (SPIDERS) DR14 cluster program value-added catalog. We list the 416 BCGs identified as part of this process, along with their stellar mass, star formation rates, and morphological properties. We identified the BCGs based on the available spectroscopic data from SPIDERS and photometric data from SDSS. We computed stellar masses and SFRs of the BCGs on the basis of SDSS, WISE, and GALEX photometry using spectral energy distribution fitting. Morphological properties for all BCGs were derived by Sersic profile fitting using the software package SIGMA in different optical bands (\textit{g,r,i}). We combined this catalog with the BCGs of galaxy groups and clusters extracted from the deeper AEGIS, CDFS, COSMOS, XMM-CFHTLS, and XMM-XXL surveys to study the stellar mass - halo mass relation using the largest sample of X-ray groups and clusters known to date. This result suggests that the mass growth of the central galaxy is controlled by the hierarchical mass growth of the host halo. We find a strong correlation between the stellar mass of BCGs and the mass of their host halos.  This relation shows no evolution since \textit{z}  $\sim$ 0.65. We measure a mean scatter of 0.21 and 0.25 for the stellar mass of BCGs in a given halo mass at low ($0.1<z < 0.3$) and high ($0.3<z<0.65$) redshifts, respectively.  We further demonstrate that the BCG mass is covariant with the richness of the host halos in the very X-ray luminous systems.  We also find evidence that part of the scatter between X-ray luminosity and richness can be reduced by considering stellar mass as an additional variable. }

\maketitle

\section{Introduction} 

The brightest cluster galaxies (BCGs) are the most luminous galaxies among the member galaxies in a cluster. These galaxies usually reside close to the optical and X-ray centers of galaxy clusters (or groups), depending on the dynamical state of the clusters. Previous studies show that BCGs do not follow the same luminosity function as satellite galaxies (\citealt{Tremaine1977}), suggesting a distinct evolutionary path resulting in a unique set of properties.  Moreover, the unique location of these luminous galaxies link their origin and evolution with the evolution of their host clusters, and therefore can provide direct information on the formation history of large-scale structures in the Universe \citep{Conroy2007}. 

Although, many pioneering studies have attempted to establish a single scenario for BCG formation \citep{Fabian1994,Cowie1977,Fabian1977,Ostriker1977, Ostriker1975,Kormendy1984,Merritt1983,Merritt1984}, recent studies have generally suggested that no single scenario can fully explain all aspects of BCGs \citep{Brough2007, Loubser2012, Jimmy2013}. For instance, BCGs
display a very diverse range of mass-to-light ratios, of sizes at a given mass, and spatial distributions of their stellar populations \citep{vonderLinden2007, Loubser2012}. 

There are many unanswered questions on the formation and evolution of BCGs. In the current paradigm of structure formation in the universe, galaxies form within their cold dark matter halos. These dark matter halos mainly form and evolve by gravity. However, the assembly of the stellar content of galaxies is regulated by the more complex physics of gas cooling, heating, and consumption by star formation as well as merger events. According to numerical simulations and semi-analytical models, BCGs form through a two-fold process. At high redshifts, the stellar mass component of BCGs is initially created through  the collapse of cooling gas or gas-rich mergers. The BCGs continue to grow essentially by dissipationless processes such as dry mergers \citep{DeLB07,Naab2009,Laporte2012}. In general, observations confirm this formation scenario; however, some studies imply that the key mechanism for the evolution of BCGs is feedback rather than merging  (e.g.,  \citealt{Ascaso2011}). 

In order to understand how the hierarchical growth of structures regulates the properties of a galaxy, we should quantify the relation between dark matter halos and the properties of the galaxies they host. Specifically, the stellar mass-halo mass relation provides powerful constraints on the galaxy formation process that any successful model should be able to account for. Moreover, the mass ratio of the stellar mass of BCGs to their host group or cluster reveals how efficiently the baryonic component has been converted into stars. Both direct and indirect methods have been employed to study this relation. Direct measurements of halo mass include X-ray observations, the Sunyaev$-$Zeldovich effect, galaxy$-$galaxy lensing, and satellite kinematics within galaxy groups and clusters (e.g., \citealt{Lin2004, Yang2007, Hansen2009, Kravtsov2004,Gozaliasl2019}). Indirect methods include halo occupation distribution (HOD) modeling \citep{Yang2003, Leauthaud2011, Leauthaud2012, Zehavi2011, Parejko2013, Guo2014}, conditional luminosity function modeling \citep{Yang2009}, and abundance matching techniques \citep{Colin1999, Kravtsov1999, Kravtsov2004, Tasitsiomi2004, Vale2004, vandenBosch2005, Conroy2006, Shankar2006, Vale2006, Conroy2009, Behroozi2010, Guo2010, Moster2010, Hearin2013, Reddick2013}.

In this work, we examine the stellar mass - halo mass relation of BCGs using a sample of X-ray groups and clusters with the widest halo mass range available to date.  To this end, we perform a search based on the photometric and spectroscopic redshifts provided by the Sloan Digital Sky Survey (SDSS DR14) to identify BCGs of clusters found in  SPectroscoic IDentification
of eROSITA Sources--COnstrain Dark Energy with X-ray clusters (SPIDERS-CODEX). We provide  properties of galaxies such as stellar mass, star formation rate (SFR), and structural parameters for this BCG sample. This catalog is combined with BCG catalogs extracted from the deeper All-wavelength Extended Groth strip International Survey (AEGIS) \citep{Erfanianfar2013}, Cosmological Evolution Survey (COSMOS) \citep{Finoguenov2007, George2011, George2013}, ECDFS survey \citep{Erfanianfar2014}, XMM-CFHTLS  \citep{Mirkazemi2015}, and XMM-XXL survey \citep{Pierre2016} to widen the halo mass range probed. The BCGs in all surveys are identified and characterized following identical methods.
In Section 3 we present the X-ray cluster catalog and optical data that we used in this work. In Section 3 we explain the method for determining the various BCG properties . In Section 4 we present the stellar mass - halo mass relation. We discuss and summarize our findings in Section 5. Throughout, we adopt a \citep{Chabrier2003} initial mass function (IMF), and a cosmology with H$_0$ = 71 km s$^{-1}$ Mpc$^{-1}$, $\Omega_m$ = 0.3, and $\Omega_{\Lambda}$= 0.7. The BCG value-added catalog (VAC) is available through the SDSS website\footnote{\url{http://www.sdss.org/dr14/data_access/vac/}}.

\section{Data}
\label{sec:data}
Our study is based upon a sample of BCGs drawn from a variety of datasets.  The primary sample of this work is that produced from the SDSS SPIDERS-CODEX sample of galaxy clusters; we  systematically and objectively identify 416 BCGs from this dataset and present a catalog of the properties of these objects in this paper. We supplement this sample with BCGs drawn from five previous surveys. In this section we describe the basic specifications of the six surveys.
\subsection{ CODEX - SPIDERS}
The SPIDERS-CODEX survey is one of the projects of the Sloan Digital Sky Survey-IV (SDSS-IV; \citealt{Blanton2017}). The  SPIDERS survey (\citealt{Clerc2016}) obtains spectra of X-ray sources  using the SDSS telescope and spectrograph (\citealt{Gunn2006,Smee2013}) covering the optical and near-infrared (NIR) at a spectral resolution of approximately 2000.  The host groups and clusters in our study are drawn from the catalog of spectroscopically confirmed X-ray detected clusters in the Data Release 14 the SDSS (DR14; \citealt{Abolfathi2018}).

Galaxy clusters were initially identified via the emission of their hot baryonic component as X-ray extended sources in ROSAT (ROentgen SATellite) and XMM-Newton observations. The initial redshift was assigned to these clusters using the red-sequence method. Spectroscopic redshifts obtained by SPIDERS (DR14) provided confirmation of the clustered nature of these objects and their redshift (up to z$\sim$0.65). The gas properties derived from X-ray observations, such as luminosity and R$_{500}$, were derived using precise cluster redshifts and these properties provide an estimate of their total mass $M_{200}$ following \citet{Leauthaud2010} and assuming a standard evolution of scaling relations, i.e., $M_{200} E_z = f (L_x E_z^{-1}),$ where $E_z = (\Omega_m (1+z)^3 +\Omega_\Lambda )^{1/2}$ (median $M_{200}$ $\sim 5 \times 10^{14}$ $M_{\odot}$) \citep{Clerc2016}. We used only those clusters with a single optical counterpart from this catalog.

\subsection{AEGIS, COSMOS, and CDFS}
A full description of the data available in the AEGIS, COSMO, and CDFS  fields is provided in $\S$2.1 of Erfanianfar et al. (2014). Briefly, all of these fields have been covered by deep {\itshape Chandra} and {\itshape XMM-Newton} observations \citep{Erfanianfar2014, Finoguenov2007, George2011, George2013}. In addition to deep X-ray data, all of these surveys take advantage of deep multiwavelength imaging data extending to the radio regime and a dense (optical) spectroscopic sampling. We used clusters and groups in these fields with an assigned identification flag equal to 1 and redshifts below $z = 0.65$. These criteria add an additional 45 (typically less massive) X-ray groups to our full sample.

\subsection{XMM-CFHTLS}
The XMM-CFHTLS catalog has been constructed through a series of short  {\itshape XMM-Newton} observations of faint RASS ( ROSAT all-sky survey ) sources in the CFHTLS  wide fields \citep{Mirkazemi2015}. We used the T0007 data release of CFHTLS and its corresponding photometric redshift catalog. The photometric redshifts were computed using the methods of \citet{Ilbert2006} and \citet{Coupon2009}. The spectroscopic redshifts were derived from a series of Hectospec observations with MMT\citep{Fabricant2005, Mirkazemi2015}. We also added the spectroscopic data from DR14 SDSS to this sample. We used in total 75 X-ray clusters from XMM-CFHTLS with 0.1$<z<$0.65.

\subsection{XMM-XXL}
\begin{figure}
\centering{\includegraphics[scale=0.5]{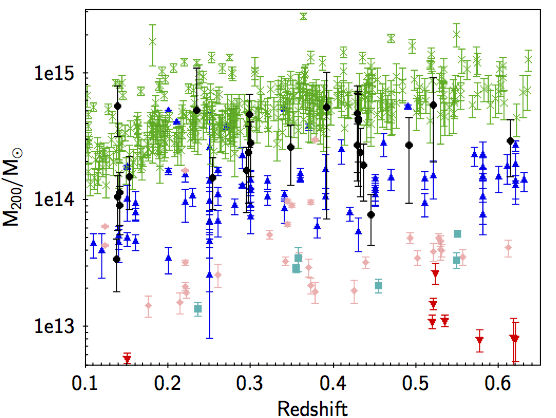}}
\caption{Cluster M$_{200}$ - redshift plane for CODEX -SPIDERS (green cross), AEGIS (cyan filled square), COSMOS (pink diamond), CDFS (point down red triangle), XMM-CFHTLS (point-up blue triangle), and XMM-XXL (black filled circle)  surveys used in this work. \label{fig:m200z}}
\end{figure}

The XXL Survey is the largest area (50 deg$^2$) XMM program to date \citep{Pierre2016} with the aim of  identifying several hundred galaxy clusters along with several tens of thousands of AGNs in the [0.5-2] keV band. We used those clusters which overlap with the W1 field of CFHTLS  to take advantage of the photometric and spectroscopic data available in this wide field. In total, this adds 25 (predominantly massive) X-ray clusters from XMM-XXL to our study. Figure~\ref{fig:m200z} shows the $M_{200}-$redshift relation for the full sample of groups and clusters used in this work.

\subsection{Data base summary}

Our sample of clusters includes 416 from SPIDERS and 110 from the other five cluster surveys. During our analysis we detected two duplications between these surveys. There was one cluster in common between CODEX and CFHTLS and one between CODEX and XMM-XXL. The X-ray properties were consistent. However, we kept those from CFHTLS and XMM-XXL due to deeper X-ray and optical data. The locations in the mass-redshift plane of the 526 unique clusters in our total sample is presented in Figure 1.  The clusters cover redshifts between 0.1 and 0.65.

\section{Method}
\subsection{BCG selection}

\begin{figure}
\includegraphics[scale=0.65]{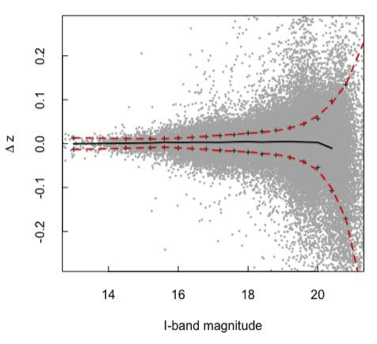}
\caption{ Shown is the relation (Photometric redshifts - Spectroscopic redshifts) / (1 + Spectroscopic redshifts) ($\Delta$z) vs. $i$-band cModel magnitude for a sample of galaxies with both spectroscopic and photometric redshifts in SDSS DR14 (gray dots). The black pluses and black solid line show the dispersion and  peak of the Gaussian fitted in each $i$-band magnitude bin, respectively. There are in total 17 bins. The red dashed lines indicate the model fitted to the dispersions. \label{fig:deltazmag}} 
\end{figure}

\begin{figure}
\includegraphics[scale=0.45]{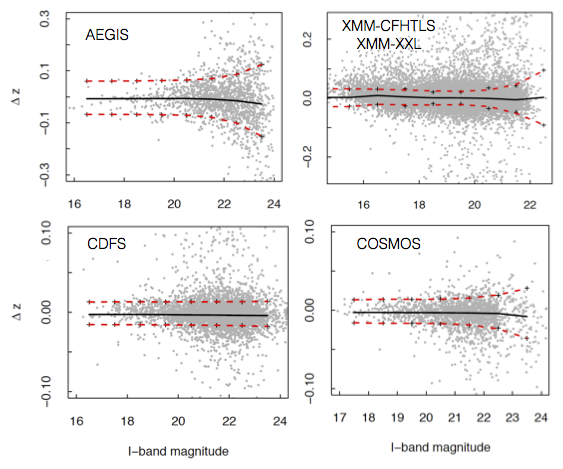}
\caption{ Same as Figure~\ref{fig:deltazmag} but for the AEGIS, COSMOS, CDFS, XMM-CFHTLS, and XMM-XXL fields. \label{fig:deltazmagall}}
\end{figure}

\begin{figure}
\centering{\includegraphics[scale=0.45]{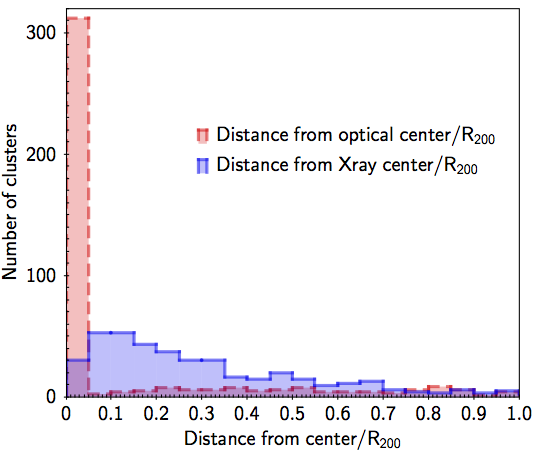}}
\caption{ Distance of the BCGs from the optical and X-ray centers normalized by R$_{200}$.\ Of the BCGs, 79\% reside in $\sim 0.05 \times r_{200}$ from optical center, and 92\%  and 84\%  of BCGs reside in 0.5$\times$R$_{200}$ from optical center and X-ray center, respectively.   \label{fig:distdist}}
\end{figure}

In many previous studies, BCGs have been identified via visual inspection. However, for a large sample of groups and clusters, it is not an efficient approach. In order to cope with this problem, we selected BCGs based on their photometric and spectroscopic properties.  For the CODEX-SPIDERS sample,  we considered all clusters from the DR14 cluster catalog that have at least one member with a spectroscopic redshift  (\citealt{Clerc2016}). According to Clerc et al. 2016, in the process of cluster identification, for some X-ray sources, two optical counterparts have been assigned. We did not consider these counterparts in this analysis to exclude any uncertainties they could impose on this work. This sample includes 439 clusters. This catalog contains both X-ray and optical centers. The X-ray center is the center of X-ray sources in the ROSAT observation. Owing to the large RASS survey point spread function (PSF) with a full width half maximum (FWHM) of $\sim$ 4 arcminutes \citep{Boese2000}, these centers have a relatively large uncertainty associated with them. The second centers in the catalog are the optical centers \citep{Rykoff2014}. The photometry and photometric redshifts of all objects within three virial radii (3$\times$r$_{200}$) from the optical center were extracted from SDSS. We only considered those galaxies that have a good photometry flag.  Photometric redshifts from DR14 SDSS were substituted for spectroscopic redshifts when available. In order to model the photometric redshifts based on the cModel$_i$ magnitude of galaxies (Figure~\ref{fig:deltazmag}), we used galaxies with spectroscopic redshift at the same distance from the cluster centers. We needed this model to know how to select member galaxies based on their photometric redshifts. We used this model, the optical center, and the redshift of the clusters to identify member galaxies. According to the galaxies' $i$-band magnitude and the model, we chose those galaxies that are within 3$\times\sigma_{model}$ as  member galaxies. \\
Appendix A demonstrates the reliability of this method in selecting member galaxies based on the photometric redshifts. Using the halo lightcone catalog from the Millennium$_{XXL}$ simulation, we show that the accuracy of BCG selection based on the photometric redshift of galaxies is$ \sim 88\%$.  In case a spectroscopic redshift has been measured, we consider that galaxy as a cluster member if its redshift falls within 0.01$\times$(1 $+ z_{\rm cluster}$). After selecting member galaxies with this method, we chose BCGs as the brightest member galaxy inside 1$\times r_{200}$ from the optical center; 81\% of BCGs have both spectroscopic and photometric redshifts. Figure~\ref{fig:distdist} presents the distance of the BCGs from the optical and X-ray centers normalized by R$_{200}$. Most of the BCGs (79\%) reside in $\sim 0.05 \times r_{200}$ from optical center. Approximately 92\%  and 84\%  of BCGs reside in 0.5$\times$R$_{200}$ from optical center and X-ray center, respectively. To produce a homogeneous BCG sample, we selected the BCGs using the same procedure as for CODEX-SPIDERS for all of five other fields (see Figure~\ref{fig:deltazmagall}). 

\subsection{Stellar mass and star formation rate}

We computed the SFR and stellar masses for the BCGs of CODEX-SPIDERS clusters using Le PHARE (PHotometric Analysis for Redshift Estimations; \citealt{Arnouts2001,Ilbert2006}), a publicly available code based on a ${\chi}^2$ template fitting procedure. We used Galaxy Evolution Explorer (GALEX) data for the ultraviolet (UV) part of the spectrum. The GALEX instrument provides UV broadband information: far-UV (FUV, 1344-1786\AA) and near-UV (NUV, 1771-2831\AA). For the optical part of the spectrum we used SDSS $u, g, r, i , z$ cModel magnitudes. We also used Wide-Field Infrared Survey Explorer (WISE) forced photometry \citep{Lang2016} for the infrared part of spectrum.\\
 We followed the procedure described in \citet{Ilbert2009, Ilbert2010}. First we adjusted the photometric zero points, as explained in  \citet{Ilbert2006}. Namely, using a ${\chi}^2$ minimization at fixed cluster redshift, we determined for each galaxy the corresponding best-fitting COSMOS templates (included in the package; see \citealt{Ilbert2006}). Dust extinction was applied to the templates using the \citet{Calzetti2000} law, with E(B-V) in the range 0-0.5 and with a step of 0.1.
We applied the systematic zero-point offsets to the observed-frame photometry and computed the SFR and stellar masses using LePHARE, following the recipe of Ilbert et al. (2010). The spectral energy distribution (SED) templates for the computation of mass and SFR were generated with the stellar population synthesis package developed by \citep[][hereafter BC03]{Bruzual2003}. We assume a universal IMF from \citet{Chabrier2003} and an exponentially declining SF history, SFR $\propto$ e$^{-t/\tau}$ (with 0.1$ < \tau <$ 30 Gyr). The SEDs were generated for a grid of 51 ages (spanning a range from 0.1 Gyr to 13.5 Gyr). We also included emission lines appropriate for the templates as described in Ilbert et al. (2010). We replaced the SFR from SED fitting with those from MPA-JHU VAC. Stellar mass estimates for the other fields in this work were derived through SED fitting by Le PHARE \citep{Ilbert2010, Erfanianfar2014, Mirkazemi2015}. Consistent stellar mass estimations are key to prevent potential biases in the analysis of our combined sample.

\subsection{Structure parameters}

The BCGs in the SPIDERS sample were modeled using GALFIT \citep{Peng2010} through the Structural Investigation of Galaxies via Model Analysis code (SIGMA), which is the host pipeline that runs GALFIT (\citealt{Kelvin2012}; see relevant sections in \citealt{Furnell2018}). An R-based pipeline software previously used in the GAMA (Galaxy And Mass Assembly) survey \citep{Driver2011}, SIGMA provides model light profiles for $\sim$$10^{5}$ GAMA galaxies in five optical (SDSS$-ugriz$) and four NIR (UKIRT$-$YJHK) passbands (see \citealt{Hill2011}). This pipeline is capable of performing a full fit, including: object extraction through Source Extractor (SExtractor; \citealt{Bertin1996}), creating a model of the field PSF (PSFEx; \citealt{Bertin2011}), estimating the local sky about an object, masking external objects and, finally, fitting a two-dimensional light profile through GALFIT. We provide fits for our BCGs in three bands (gri) for three models: a de Vaucouleurs profile (n = 4), a free Sersic profile (1 $<$ n $<$ 20), and a free Sersic + fixed exponential (n = 1) dual-component model (\citealt{Furnell2018}). Table 1 lists the columns of the CODEX-SPIDES VAC; C1 and C2 refer to the first and second components to a model, for the S+X model; C1 is the Sersic bulge and C2 is the exponential halo.

    \begin{table*}
    \begin{threeparttable}
    \caption{CODEX-SPIDERS DR14 BCGs value-added catalog}
    \label{tb:cor:cha}
    \centering
    \small
  
\begin{tabular*}{\textwidth}{ll}
    \hline
     \toprule
     Name of columns & description \\
      \hline
       \midrule
1. CLUS$\_$ID &   The SPIDERS/CODEX identification number {i}$\_${nnnnn}  \\
2. CLUZSPEC &   Galaxy cluster redshift     \\
3. RA$\_$BCG &  BCG right ascension (J2000)  \\
4. DEC$\_$BCG & BCG declination (J2000)   \\
5. Mass$\_$MEDIAN  &    log(Stellar Mass)     \\
6. SFR$\_$MEDIAN  &log(SFR)\\
7. flag$\_$SFR$\_$MPA$\_$JHU & =1 if from MPA$\_$JHU VAC    \\
8. GAL$\_$SDSS$\_$(g$|$r$|$i$|$z)$\_$modS$\_$(C1$|$C2)$\_$CHI2NU & Reduced ${\chi}^2$ for single-Sersic fit in the (g$|$r$|$i$|$z) band  \\
9. GAL$\_$SDSS$\_$(g$|$r$|$i$|$z)$\_$modS$\_$(C1$|$C2)$\_$MAG &Primary object magnitude for single-Sersic fit in the(g$|$r$|$i$|$z) band \\
10. GAL$\_$SDSS$\_$(g$|$r$|$i$|$z)$\_$modS$\_$(C1$|$C2)$\_$RE & Primary object effective radius for single-Sersic fit in the (g$|$r$|$i$|$z) band\\
11. GAL$\_$SDSS$\_$(g$|$r$|$i$|$z)$\_$modS$\_$(C1$|$C2)$\_$N & Primary object Sersic index for single-Sersic fit in the (g$|$r$|$i$|$z) band\\
12. GAL$\_$SDSS$\_$(g$|$r$|$i$|$z)$\_$modS$\_$(C1$|$C2)$\_$AR & Primary object axis ratio for single-Sersic fit in the (g$|$r$|$i$|$z) band\\
13. GAL$\_$SDSS$\_$(g$|$r$|$i$|$z)$\_$modS$\_$(C1$|$C2)$\_$PA & Primary object position angle for single-Sersic fit in the (g$|$r$|$i$|$z) band\\
14. GAL$\_$SDSS$\_$(g$|$r$|$i$|$z)$\_$modS$\_$(C1$|$C2)$\_$MAG$\_$ERR & Error on magnitude for single-Sersic fit in the (g$|$r$|$i$|$z) band\\
15. GAL$\_$SDSS$\_$(g$|$r$|$i$|$z)$\_$modS$\_$(C1$|$C2)$\_$RE$\_$ERR & Error on effective radius of primary object for single-Sersic fit in the (g$|$r$|$i$|$z) band\\
16. GAL$\_$SDSS$\_$(g$|$r$|$i$|$z)$\_$modS$\_$(C1$|$C2)$\_$NvERR & Error on Sersic index of primary object for single-Sersic fit in the (g$|$r$|$i$|$z) band\\
17. GAL$\_$SDSS$\_$(g$|$r$|$i$|$z)$\_$modS$\_$(C1$|$C2)$\_$AR$\_$ERR & Error on axis ratio of primary object for single-Sersic fit in the (g$|$r$|$i$|$z) band\\
18. GAL$\_$SDSS$\_$(g$|$r$|$i$|$z)$\_$modS$\_$(C1$|$C2)$\_$PA$\_$ERR & Error on position angle of primary object for single-Sersic fit in the (g$|$r$|$i$|$z) band\\

       \hline
    \bottomrule
        
     \end{tabular*}

  \end{threeparttable}
\end{table*}

  \begin{table}
  \begin{threeparttable}
    \caption{Best-fit power-law parameters for the stellar mass -- halo mass relation}
        \label{table:shmr}
     \begin{tabular}{lll}
       \hline
        \toprule
        
        Relation &slope &normalization \\
         \hline
      \midrule
                 &     \multicolumn{1}{c}{0.1$\le z \le$0.3}     \\[0.1cm]
        M$_{*,BCG}$-M$_{200}$&  $0.41\pm0.04$ & $5.59\pm0.54$ \\[0.1cm]
        M$_{*,BCG}$/M$_{200}$-M$_{200}$ & $-0.58\pm0.038$ & $5.467\pm0.52$ \\[0.2cm]
          &     \multicolumn{1}{c}{0.3$<$z$\le$0.65}     \\[0.1cm]
        M$_{*,BCG}$-M$_{200}$&  $0.31\pm0.02$ & $7.00\pm0.37$ \\[0.1cm]
        M$_{*,BCG}$/M$_{200}$-M$_{200}$ &  $-0.79\pm0.03$ &$8.383\pm0.50$ \\[0.1cm]
        \bottomrule
        \hline
     \end{tabular}
    \begin{tablenotes}
\item{     \small
            The relations are fit by the power law y = mx + c, where
    x=log$_{10}$(M$_{200}$/M$_{\odot}$)and y is log$_{10}$(M$_{*,BCG}$/M$_{\odot}$)(or log$_{10}$M$_{*,BCG}$/M$_{200}$)}
    \end{tablenotes}
  \end{threeparttable}
\end{table}

  \begin{table}
  \begin{threeparttable}
    \caption{Mean of stellar mass of BCGs in a given halo mass}

     \begin{tabular}{ c c c c}
       \hline
        \toprule
        
        Halo mass & Mean(M$_{*,BCG}$) & $\sigma_{LogM_*}$ & number of clusters \\
         \hline
      \midrule
                 &     \multicolumn{2}{c}{0.1$\le z \le$0.3}     \\[0.1cm]
        13.2&11.04&0.20 &15 \\[0.1cm]
        13.8&11.2&0.20&15 \\[0.1cm]
        14.2&11.46 &0.24&57\\[0.1cm]
        14.5&11.61 &0.20&78 \\[0.1cm]
        14.8 &11.66 &0.20&79\\[0.2cm]
          &     \multicolumn{2}{c}{0.3$<z\le$0.65}     \\[0.1cm]
        13.2&  11.16 & 0.28 & 18 \\[0.1cm]
        13.7&  11.19 & 0.21 &16\\[0.1cm]
        14&  11.40 & 0.25 & 15\\[0.1cm]
        14.2&  11.46 & 0.25 &19\\[0.1cm]
        14.5&  11.57 & 0.27 &21\\[0.1cm]
         14.7&  11.58 & 0.25 &85\\[0.1cm]
         14.9&  11.58 & 0.24&76 \\[0.2cm]
        \bottomrule
        \hline
     \end{tabular}
     \end{threeparttable}
\end{table}
\begin{figure}
\centering{ \centering

\centering{\includegraphics[scale=0.4]{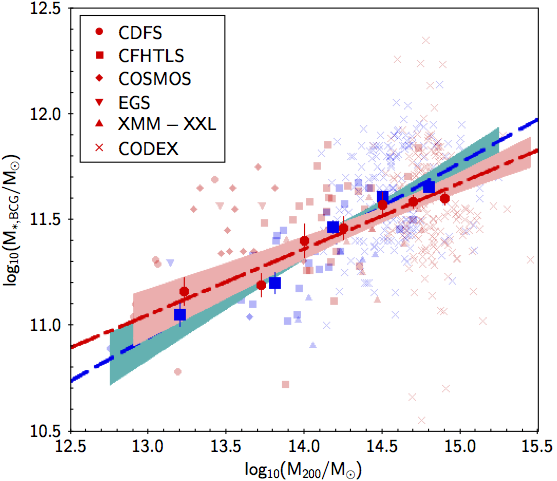}}}  
\caption{ Stellar masses vs. halo masses of the BCGs in our sample. Different symbols represent different surveys. The blue shaded symbols show BCGs with 0.1$\le$z$\le$0.3 and the red shaded symbols show BCGs with 0.3$<$z$\le$0.65. The blue dashed and the red dash-dotted lines show the best power-law fits for the low and high redshifts, respectively. The blue and red shaded area around the fitted lines represent 95\% confidence levels of the fits. The blue squares and red dots with error bars present the mean of stellar mass of BCGs at a given halo mass. The error bars are their standard errors. The Pearson correlation coefficient is 0.7 for low redshift BCGs and 0.4 for high redshift BCGs. \label{fig:mm}}
\end{figure}

\begin{figure}
\centering{ \centering
\centering{\includegraphics[scale=0.5]{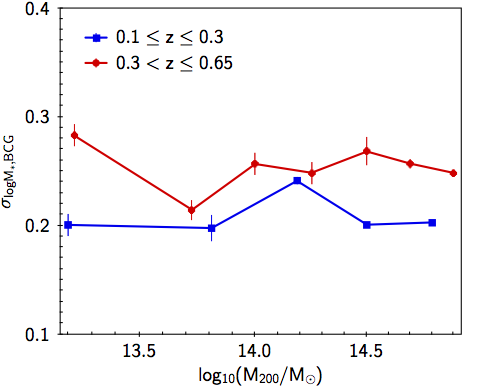}}}  
\caption{ Scatter in the stellar mass of BCGs at a given halo mass in low and high redshift bins. The scatter is the quadratic sum of intrinsic scatter and measurement scatter of stellar masses. \label{fig:scatter}}
\end{figure}

\begin{figure}
\centering{ \centering
\centering{\includegraphics[scale=0.6]{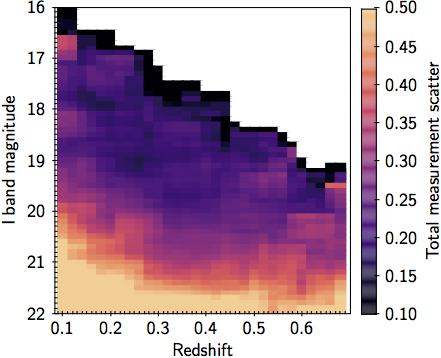}}}
\caption{ Total measurement scatter in estimating stellar mass in redshift-magnitude space. \label{fig:scattererr}}
\end{figure}

\begin{figure*}
\centering{
\centering{\includegraphics[scale=0.55]{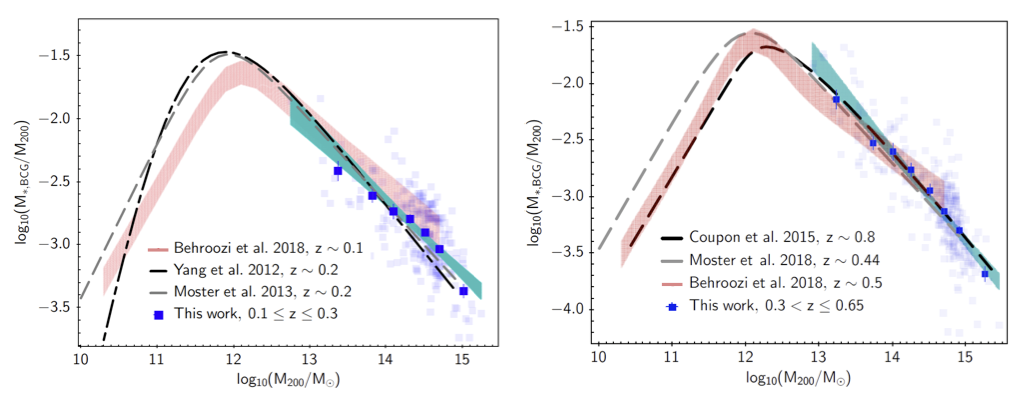}}}  
\caption{ Stellar-to-halo mass ratio vs. host halo mass of BCGs. The shaded blue areas and the blue squares show the 95\% confidence level of the fitted model and the mean of the stellar-to-halo mass ratio of BCGs at a given halo mass, respectively. Their error bars represent the standard error in stellar mass fraction including intrinsic and measurement uncertainties. The red shaded area is the same relation from \citet{Behroozi2010} at $z\sim$0.1 and $z\sim$0.5, respectively. The dash-dotted black line in the left panel shows the Yang et al. (2009) relation at z$\sim$0.1. The gray dashed lines show the \citet{Moster2018} relation. \label{fig:SMHM}}
\end{figure*}

\section{Stellar mass - halo mass relation}
In this section, we explore the connection between the stellar mass of BCGs and the halo mass of their host groups or clusters. For this purpose, we divide our sample into two redshift bins (low redshift : 0.1$\le$z$\le$0.3 and high redshift : 0.3$<$z$\le$0.65). This is also done in order to take into account different sampling of X-ray luminosity with redshift. First, we compare the stellar mass of BCGs with the halo mass of their host groups and clusters (see Figure \ref{fig:mm}). The red and blue shaded points show the individual BCGs and the blue and red lines represent the fitted power-law model. The shaded areas indicate the corresponding 95\% confidence levels. The blue and red points with error bars present the geometric mean of BCG stellar masses for a given halo mass. The stellar mass of BCGs and their host halo masses correlate significantly, with Pearson correlation coefficients of 0.7 and 0.4 in the low and high redshift bins, respectively. The two scaling relations agree with each other within their uncertainties. We do not observe any notable redshift evolution in the relation between the central galaxy stellar mass and the host halo mass since z$\sim$0.65. The best power-law fitted model is listed in table \ref{table:shmr}. We also report the mean and the corresponding scatters in table 3. The bins of halo masses are chosen such that each bin includes at least 15 clusters.

Figure \ref{fig:scatter} shows the dispersion in the stellar mass of BCGs at a given halo mass ($\sigma_{logM_*}$). This scatter is one of the most fundamental aspects of the relationship between the galaxies and their host halos \citep{Wechsler2018}. In abundance matching models, we can derive the same stellar mass function from different choices for scatter, but the predicted spatial distribution of galaxies would be different. This directly impacts the clustering of galaxies. As massive halos ($M_h$ $> 10^{12} M_{\odot}$) get more clustered, thus galaxies at high masses have the highest sensitivity to this scatter \citep{Wechsler2018}. We derive a value of 0.21 and 0.25 for  the mean scatter of the stellar mass at a given halo mass at low and high redshifts, respectively. The slightly higher scatter at high redshift is consistent with the larger uncertainties of stellar mass estimates. This is the first time that this scatter is quantified directly using a sample of X-ray groups and clusters over a halo mass range spanning 13 $<$log(M$_{200}/M_{\odot})$$<$ 15.4. Previously, \citet{Kravtsov2018}, using a small sample of local clusters, derived a value of ($\sigma_{logM_*}$)$\sim$0.19. Moreover, \citet{Zu2015} obtained a constraint on the scatter at $M_h$$\sim$$10^{14}$ $M{_\odot}$ ($\sigma_{logM_*}$ = 0.18$\pm$0.01) using a combination of galaxy clustering and galaxy lensing of $z $= 0 SDSS galaxies. Current predictions from the semi-analytic and empirical models have consistent scatter in the halo mass range considered in this work (\citealt{Henriques2015, Lu2014, Somerville2012,Behroozi2018}, see Figure 8 in \citealt{Wechsler2018}). However, some of  hydrodynamical simulations predict a scatter below 0.2 dex at the high-mass end (\citealt{Khandai2015, McAlpine2016, Pillepich2018}).  We note that the observed $\sigma_{logM_*}$ constitutes the quadratic sum of intrinsic scatter and measurement uncertainties on stellar masses.  A proper understanding of the measurement error in stellar mass is thus required to infer the intrinsic scatter.  In this work, we consider as sources of the measurement dispersion both the propagated uncertainties in the photometry and the photometric redshifts based thereupon.  \\

Figure \ref{fig:scattererr} presents the measurement scatter of stellar mass in redshift-magnitude space for the CODEX sample. We derive the scatter by propagating the errors in magnitudes based on the magnitude uncertainties in SDSS. Moreover, we include uncertainty induced by the lack of NIR  data.  We also consider an intrinsic error in stellar mass
computation using the SED fitting method, accounting for uncertainties associated with, for example, the adopted functional form for the star formation histories and reddening law ($\sim$ 0.14 dex see \citealt{Ilbert2010}). Figure \ref{fig:scattererr} illustrates that the scatter induced by stellar mass measurement is $\sim$ 0.19 dex for galaxies with $i<$21 and 0.1 $\le z \le$ 0.65. Taking advantage of deeper photometry and often the existence of NIR data, we would expect a lower scatter in our stellar mass measurement for our other surveys \citep{Erfanianfar2014}. However, since we are working with BCGs, this difference is not significant and does not cause differences in the dispersion of stellar mass of BCGs versus the mass of their host halos for different surveys (See Fig. \ref{fig:mm}).\\

We investigate the relation between the stellar mass to host halo mass ratio of BCGs as a function of halo masses (SHMR; Figure \ref{fig:SMHM}). For comparison, we also show inferences from the literature by \citet{Behroozi2018}, i.e., SHMR using empirical models based on evolving galaxies within their dark matter halo histories constrained by galaxy clustering and galaxy-galaxy lensing; \citet{Moster2018}, i.e., the parameterized SHMR inferred from abundance matching; \citet{Kravtsov2012}, i.e., SHMR from observed X-ray clusters at low redshift; \citet{Yang2012}, i.e., SHMR from conditional luminosity function modeling; and \citet{Coupon2015}, SHMR from HOD modeling). The large blue squares indicate the mean of the stellar masses of BCGs at a given halo mass.  A comparison with \citet{Behroozi2018} and \citet{Moster2018} agrees well within the error bars at both low and high redshift. We report the best power-law fitted model in table \ref{table:shmr}.

\section{Stellar mass-richness correlation}

\begin{figure}
\centering{ \centering
\centering{\includegraphics[scale=0.45]{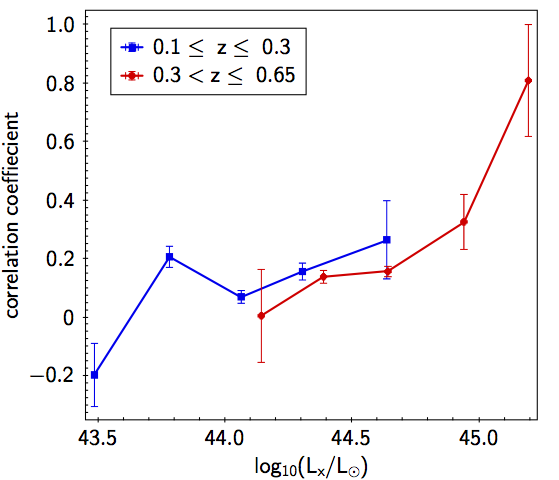}}}  
\caption{ Correlation coefficient for stellar mass of BCGs and the richness of the host halos in a given X-ray luminosity. The error bars are derived using jackknife resampling\label{fig:corr}}
\end{figure}

As a final test, we investigated the Pearson correlation (normalized covariance) between the richness and the stellar mass of BCGs at a given X-ray luminosity. Figure \ref{fig:corr} shows this correlation at low and high redshifts. We only use the CODEX sample, for which measurements of richness are available from redMapper \citep{Rykoff2014}. At least 15 clusters exist in each bin of X-ray luminosity. The error bars are estimated through the Jacknife procedure \citep{Efron1982}. There is a slightly positive richness-BCG mass correlation starting at logL$_X>$ 44.5. This correlation becomes stronger at the high luminosity end of our sample. Previously, \citet{Furnell2018}, using a sample of massive clusters with 0.05 $\le$ z $\le$ 0.3, found a weak positive correlation between stellar mass of BCGs and richness of clusters in a given X-ray luminosity. The positive correlation between the stellar mass of  BCGs and the richness of clusters implies that the growth of the BCGs and the growth of the host clusters are intimately related.  Indeed, the deviation in richness is often associated with higher BCG mass. This positive correlation may also provide clues to the origin of the dispersion of the richness and X-ray luminosities in extremely luminous halos. 

\section{Summary and discussion}

In this paper, we derived the properties of BCGs of a large sample of X-ray clusters in the CODEX, COSMOS, AEGIS, CFHTLS-XMM, and XMM-XXL surveys with spectroscopic and photometric redshifts. We investigated the distance of the BCGs from the X-ray and optical centers in the CODEX sample. We find BCGs positions agree well with the optical centers ( Figure \ref{fig:distdist}). The BCGs of CODEX clusters and their properties are presented as a VAC of the SDSS DR14 \citep{Abolfathi2018}. \\

By estimating the stellar masses of BCGs in a consistent manner, we constructed the largest current BCG sample from X-ray halos to study the stellar mass $-$ halo mass relation.  Investigating the galaxy -- halo connection sheds light on the physical processes regulating galaxy formation. Figure \ref{fig:mm} demonstrates that there is a strong correlation between the stellar mass and the halo masses of BCGs. We observe no evolution in this relation since z$\sim$0.65. This result suggests that the mass growth of the central galaxy is controlled by the hierarchical mass growth of the host halo. Thus, in contrast to satellite galaxies, merger events likely play a main role in the mass accretion history of central galaxies with respect to the star formation activity. This interpretation is in agreement with \citet{Behroozi2018}, who have reported that the fraction of galaxy buildup due to mergers is a strongly increasing function of mass and nearly all of dwarf galaxy buildup is due to in situ star formation and most of present-day massive galaxy buildup due to mergers. \\

We also investigated the dispersion around the mean of stellar mass of BCGs at a given halo mass. This scatter is one of the most fundamental aspects of the relationship between galaxies and their host halos. In abundance matching models, we can derive the same stellar mass function from different choices for scatter, but the predicted spatial distribution of galaxies would be different. This distribution impacts the clustering of galaxies directly. As massive halos ( $M_h$ $> 10^{12} M_{\odot}$) get more clustered, thus galaxies at high masses have the highest sensitivity to this scatter \citep{Wechsler2018}. We find an observed constant scatter of 0.21 dex at low redshifts and 0.25 dex at high redshifts. This measurement is consistent with the current predictions from hydrodynamical simulations and with some of the empirical models (\citealt{Henriques2015, Lu2014, Somerville2012,Behroozi2018}, also see Figure 8 in \citealt{Wechsler2018}). \\

As a further step, we quantified the errors in the measurement of stellar masses (Figure \ref{fig:scattererr} ). The upper limit mean scatter induced by the stellar mass measurement is $\sim$ 0.19 dex for galaxies with $i_{mag}<$21 and 0.1 $\le$ z $\le$ 0.65 in the CODEX sample, and less for the other, deeper surveys used in this work.  In order to investigate further the connection of the BCGs and their host halos, we investigated the SHMR (Figure \ref{fig:SMHM}). We find a strong agreement with \cite{Behroozi2018}, \cite{Moster2018}, \cite{Yang2012} and \cite{Coupon2015} at both low and high redshifts.\\

Finally, we show that the BCG mass is covariant with the number of galaxies hosted by the group or cluster. This positive covariance suggests a connection between the growth of the BCGs and their host halos. The strength of this correlation vanishes for lower luminous systems. This transition in observed covariance indicates a transition in the physical processes that determine the mass of a BCG. Feedback in massive BCGs could be responsible for the loss of this covariance and slightly decouple the growth of the BCGs from the growth of the cluster. We will investigate this correlation in more detail with a larger sample of clusters in RASS-DECaLS in a future study (Erfanianfar et al. in prep.).

\label{sec:edensity}

\appendix
\section{ Evaluation of the BCG selection method}

In order to evaluate the accuracy of our BCG selection in cluster surveys with low spectroscopic incompleteness, we used a halo lightcone catalog from Millennium$_{XXL}$ simulation \citep{Smith2017} . Smith et al. used the Monte Carlo method to assign galaxies randomly based on their luminosities to dark matter halos, following a HOD \citep{Smith2017}. We used 300 deg$^2$ of this catalog, which encompass 1448 galaxy clusters with virial mass $>$10$^{14}M_{\odot}$ and 0.1$\le z \le$0.65.
The photometric redshifts are randomly assigned to the galaxies using a Gaussian distribution with standard deviation given by the relation between photometric redshift dispersion and magnitude in the SDSS survey. We selected the member galaxies within one virial radius in projection on the sky and within +/- 3 times the photometric redshift error around the cluster redshift in redshift space. Since the true BCG is known from the full three-dimensional simulation information, we can assess whether the true BCG is correctly selected. We repeated the procedure of assigning noise to the photometry and finding the BCG 100 times. We find that the BCGs are selected with an accuracy of 90\%. As a further step, we also take into account the errors on the luminosity of BCGs in our analysis. In this exercise, we randomly assigned magnitude to the galaxies using Gaussian distribution with standard deviation given by the relation between magnitude error and magnitude in SDSS. We repeated the same procedure as before considering photometric redshift error and luminosity error at the same time to asses the accuracy of selecting of BCGs. This exercise demonstrates that taking into account both uncertainties, the accuracy of selection BCG decreases to 88 \%.
However, in this test, the member galaxies are selected based on photometric redshifts and the result of this test is the lower limits for our observational work.

\section*{Acknowledgments}

This paper is partially based on SDSS data. Funding for the Sloan Digital Sky Survey IV has been provided by the Alfred P. Sloan Foundation, the U.S. Department of Energy Office of Science, and the Participating Institutions. SDSS acknowledges support and resources from the Center for High-Performance Computing at the University of Utah. The SDSS website is www.sdss.org. SDSS is managed by the Astrophysical Research Consortium for the Participating Institutions of the SDSS Collaboration including the Brazilian Participation Group, the Carnegie Institution for Science, Carnegie Mellon University, the Chilean Participation Group, the French Participation Group, Harvard-Smithsonian Center for Astrophysics, Instituto de Astrof\'isica de Canarias, The Johns Hopkins University, Kavli Institute for the Physics and Mathematics of the Universe (IPMU) / University of Tokyo, the Korean Participation Group, Lawrence Berkeley National Laboratory, Leibniz Institut fur Astrophysik Potsdam (AIP), Max-Planck-Institut fur Astronomie (MPIA Heidelberg), Max-Planck-Institut fur Astrophysik (MPA Garching), Max-Planck-Institut fur Extraterrestrische Physik (MPE), National Astronomical Observatories of China, New Mexico State University, New York University, University of Notre Dame, Observatorio Nacional / MCTI, The Ohio State University, Pennsylvania State University, Shanghai Astronomical Observatory, United Kingdom Participation Group, Universidad Nacional Autonoma de Mexico, University of Arizona, University of Colorado Boulder, University of Oxford, University of Portsmouth, University of Utah, University of Virginia, University of Washington, University of Wisconsin, Vanderbilt University, and Yale University. This work also got support from the Excellence Cluster Universe with its data center C2PAP.

\bibliographystyle{aa} 
\bibliography{lib.bib}

\label{lastpage}
\end{document}